\documentclass[12pt]{article}
\usepackage{amssymb}
\usepackage{amsmath}
\usepackage{setspace}
\usepackage{url}
\usepackage[dvips,letterpaper,margin=2cm,includefoot]{geometry}

\usepackage{graphicx}
\usepackage{dcolumn}
\usepackage{bm}
\usepackage{subfig}
\usepackage[percent]{overpic}

\makeatletter
\newcommand\ackname{Acknowledgements}
\if@titlepage
  \newenvironment{acknowledgements}{%
      \titlepage
      \null\vfil
      \@beginparpenalty\@lowpenalty
      \begin{center}%
        \bfseries \ackname
        \@endparpenalty\@M
      \end{center}}%
     {\par\vfil\null\endtitlepage}
\else
  \newenvironment{acknowledgements}{%
      \if@twocolumn
        \section*{\abstractname}%
      \else
        \small
        \begin{center}%
          {\bfseries \ackname\vspace{-.5em}\vspace{\z@}}%
        \end{center}%
        \quotation
      \fi}
      {\if@twocolumn\else\endquotation\fi}
\fi
\makeatother

\begin{document}


\title{The Metabolism and Growth of Web Forums}

\author{Lingfei Wu\footnote{1. Baidu Inc., Beijing, China 2. Department of Media and Communication, City University of Hong Kong.\texttt{wlf850927@gmail.com}} \and
        Jiang Zhang\footnote{Corresponding Author. School of System Science, Beijing Normal University. \texttt{zhangjiang@bnu.edu.cn}} \and
        Min Zhao\footnote{Baidu Inc., Beijing, China \texttt{zhaomin@baidu.com}}
        }

\date{\today}

\maketitle

\begin{abstract}
We view web forums as virtual living organisms feeding on user's attention and investigate how these organisms grow at the expense of collective attention. We find that the ``body mass" ($PV$) and ``energy consumption" ($UV$) of the studied forums exhibits the allometric growth property, i.e., $PV_t \sim UV_t ^ \theta$. This implies that within a forum, the network transporting attention flow between threads has a structure invariant of time, despite of the continuously changing of the nodes (threads) and edges (clickstreams). The observed time-invariant topology allows us to explain the dynamics of networks by the behavior of threads. In particular, we describe the clickstream dissipation on threads using the function $D_i \sim T_i ^ \gamma$, in which $T_i$ is the clickstreams to node $i$ and $D_i$ is the clickstream dissipated from $i$. It turns out that $\gamma$, an indicator for dissipation efficiency, is negatively correlated with $\theta$ and $1/\gamma$ sets the lower boundary for $\theta$. Our findings have practical consequences. For example, $\theta$ can be used as a measure of the ``stickiness" of forums, because it quantifies the stable ability of forums to convert $UV$ into $PV$, i.e., to remain users ``lock-in" the forum. Meanwhile, the correlation between $\gamma$ and $\theta$ provides a convenient method to evaluate the `stickiness" of forums. Finally, we discuss an optimized ``body mass" of forums at around $10^5$ that minimizes $\gamma$ and maximizes $\theta$.
\end{abstract}

\newpage


\section{Introduction}

A Web forum is an online discussion site allowing its members to exchange opinions by posting and replying threads. Although forum is one of the oldest Internet services, its user-generated-content nature allows it to remain as one of the most frequently used services in the era of Web 2.0 \cite{o2007web,top2012500}.

The importance of Web forums has motivated many studies on user's interactions on this platform, such as detecting online opinion leaders \cite{bodendorf2009detecting}, analyzing political debates \cite{cammaerts2005online}, or identifying interest-groups \cite{zhang2007expertise,abbasi2005applying}. But these studies on forum usage usually focus on posting behavior and not browsing behavior. A major reason is that the browsing records are generally not publicly accessible. However, as a large proportion of forum users are ``silent" users who only read threads and do not give any comment \cite{benevenuto2009characterizing, yu2010analyzing}, the analysis of forum usage based on posting dynamics has strong limitations.

In this paper, we get access to the historical data of Baidu Tieba, a very large Chinese Web Forum system, and investigate the browsing activities on $30,000$ forums. The size (average daily page views) of the studied forums vary from hundreds to millions. Different from previous studies that try to understand how users use forums, we propose to study how forums ``consume" users. Specifically, we view forums as virtual living organisms that grows at the expense of user's attention. From this perspective, we can talk about the ``metabolism" of forums, which describes how the attention of users are ``absorbed" and ``dissipated" by forums. Inspired by the metabolic theory of ecology \cite{west1997general,brown2004toward,zhang2010scaling}, we observe the relation between the number of page views ($PV_t$) and the number of unique visitors ($UV_t$) in the growth of forums, which are understood as the ``body mass" and ``energy consumption" of these virtual organisms, respectively. It turns out that the vast majority of the studied forums satisfies the allometric growth pattern $PV_t \sim UV_t ^ \theta$. In other words, the growth exponent $\theta = d(Log(PV_t))/d(Log(UV_t))$ keeps unchanged over time. We suggest that $\theta$ can be use to measure the ``stickiness" of forums, as an alternative to the average surfing length $L_t = PV_t/UV_t$ \cite{bucklin2002choice}. Because both of $\theta$ and $L_t$ reflects the ability of forums to remain users ``lock-in", but the former is a constant over time, whereas the latter is not.

To probe into the origins of the observed scaling relationship between $PV$ and $UV$, we compare the flow of collective attention on forums with the flow of energy in food webs and construct clickstream networks. In these clickstream networks, the nodes are threads and the edges are user's switching between threads. We propose the conservation of attention (clickstream) as a principle that constrains the evolution of clickstream networks. We find that according to this principle, the two quantities of interest, $PV$ and $UV$, can be defined both on the network level and on the node level. On the network level, $PV$ is the total weights of edges and $UV$ is the dissipated clickstreams of the entire network; on the node level, $PV$ is sum of the clickstreams to node $i$ ($T_i$) over all nodes and $UV$ is the sum of the dissipated clickstreams from $i$ ($D_{i}$) over all nodes. The equivalence of these two versions of definitions allows us to explain the dynamics of network by the behavior of nodes. In particular, we describe the dissipation of clickstreams on nodes (threads) using the scaling function $D_i \sim T_i^ \gamma$ \cite{stravskraba1999ecosystems,zhang2013allometry}. And it turns out that $\gamma$, a quantity reflecting the dissipation efficiency of networks, is negatively correlated with $\theta$. An naive analysis shows that $1/\gamma$ sets the lower boundary for $\theta$ when $\gamma > 1$. At the end of our study, we discuss an optimized ``body mass" of forums that minimizes $\gamma$ and maximizes $\theta$ at around $10^5$ daily $PV$s.

The findings of the current study not only confirm the connection between growth and topology  in complex systems \cite{west1997general,west1999fourth,banavar1999size,garlaschelli2003universal}, but also have applied meanings. For example, the observed universal relationship between ``body mass" and ``energy consumption" will help webmasters to benchmark and monitor the growth of online communities. Meanwhile, the technique to predict the long-term behavior of forums by analyzing the random snapshots of clickstream networks may contribute to many areas of the Web development, such as click prediction \cite{cheng2010personalized} and interest group recommendation \cite{fu2000mining}. In particular, $\theta$ that describes the ``stickiness" of forums can be integrated as a novel feature into the recommendation of interest-groups \cite{fu2000mining}. Last but not least, we suggest that the presented clickstream network analysis actually provides a very general framework for studying user's browsing behavior in various online systems. In applying this analysis to other types of online social systems, one simply replaces the nodes (which are threads in the currently studied networks) with other information resources accordingly, such as news, tags, videos, etc.

\section{Related work}

\subsection{\label{sec:2.1}The flow of collective attention online}

At every moment, a large number of users are ``hopping" between information resources by clicking web pages sequentially, generating a large amount of attention flow both within and across websites \cite{huberman2009social}. Cooley and other scholars use the term ``clickstream" to describe this flow of attention and suggest that the analysis of clickstream data reveals the hidden patterns of Web usage \cite{cooley2000web}. Since Huberman et al. \cite{huberman1998strong}, individual clickstreams has been extensively studied, including the correlation between surfing length and duration time \cite{johnson2003cognitive}, the effect of surfing length on user's log-off probability \cite{bucklin2003model}, and also the distribution of surfing lengths in social networks \cite{benevenuto2009characterizing}. A study that is particularly relevant to the current work is \cite{bucklin2002choice}, which proposes to use the average surfing length $L$ to describe the website ``stickiness", which reflects the ability of a site to keep visitors ``lock-in". The common limitation of these studies is that the focus of research is always independent, individual behavior, whereas how users interacts with each other in collective surfing remains unknown.

With the development of network science, there is a rising trend to integrate clickstream studies and network theories into clickstream network analysis. In clickstream networks, the nodes are information resources and the edges are the clickstreams, i.e., the collective navigation of users between resources \cite{bollen2009clickstream}. Different from individual clickstreams, clickstream networks include the rich interactions between users via information resources, thus provides novel interpretations to some online phenomena that have been extensively studied \cite{huberman2009crowdsourcing}. For example, the surge and decay of news in the public domain is always view as a consequence of the information diffusion among users \cite{lerman2010information}. But from the perspective of clickstream networks, it can also be understood in the ``reversed way" as the transmission of user's attention between news \cite{wu2007novelty}. As a general framework, clickstream network has been applied to model various online activities, such as paper reading \cite{bollen2009clickstream}, photo tagging \cite{cattuto2007semiotic} and video watching \cite{wu2009feedback}.

\subsection{\label{sec:2.2}The metabolic theory and the allometric growth of websites}

The metabolic theory of ecology \cite{west1997general,brown2004toward,zhang2010scaling} suggests that the metabolic rate, the rate at which living organisms take up, transform, and expend energy and materials, is the most fundamental biological rate in ecology \cite{brown2004toward}. As a result, from individual biomass production to population growth, many observed patterns can be described by scaling functions with exponents that is the multiples of $1/4$. For example, the ``Kleiber's law", which is the core of the metabolic theory, posits that for the majority of mammals, their energy consumption scales to the $3/4$ power of their body mass \cite{west1997general}. Since its proposal, the metabolic theory has been extended greatly. For example, Garlaschelli et al. applied it to study food webs \cite{garlaschelli2003universal,zhang2010scaling} and Bettencourt et al. use it to explain the scaling of urban cities \cite{bettencourt2007growth}. In these studies, a frequently used term is ``allometric growth". It refers to the power-law relationships between two variables in the growth of complex systems, whose exponents can be either larger or smaller than $1$. When the exponent is larger than $1$, it is called ``superlinear scaling", otherwise, it is called ``sublinear scaling" \cite{bettencourt2007growth}.

In fact, the temporal power-law relationships are also observed widely in the virtual world, even though scholars do not use the term ``allometric growth". For example, Cattuto et al. find that in the development of online tagging systems, the total number of tags scales to the length of tag vocabulary with an exponent approaches $1.4$ \cite{cattuto2009collective}. Tessone et al. find that in social programming projects, the number of library citation relationships scales to the number of libraries with an exponent between $1.25$ and $2$ \cite{tessone2011sustainable}. Our previous studies discover a scaling relationship between the active population and the generated activities with an exponent between $1.18$ and $1.5$ \cite{wu2011accelerating}. Similar patterns are also observed in other online collective behaviors such as game playing \cite{henderson2001modelling} and email sending/receiving \cite{leskovec2007graph}. However, despite the fact that these studies have investigated various temporal scaling relationships, there is still a lack of research to examine the online version of ``Kleiber's law", that is, the power-law relationship between ``body mass" and ``energy consumption" of websites. And this is the major concern of the current study. We believe that the online version of ``Kleiber's law", once confirmed, will motive many studies that apply the metabolic theory to explain the various behaviors of websites, such as info-mass production (we use this term to refer to the increase of the user-generated content, which can be viewed as a online counterpart of the biomass production), ontogenetic growth, survival and mortality, etc.

\section{Methods}

\subsection{\label{sec:3.1} The ``body mass" and ``energy consumption" of forums}

If we understand online communities as virtual living organisms that feed on user's attention, a particularly interesting questions would be, what are the counterparts of ``body mass" and ``energy consumption" of these virtual entities ? \cite{banavar1999size} provides an flow network model to explain Kleiber's law by arguing that living organisms are, by their very nature, flow systems that transport waters and nutrient. According to this model, ``body mass" is the total amount of flow circulating within the system and ``energy consumption" is the amount of flow a system receives from the environment or dissipate to the environment (which should be equal each other). By applying this model to clickstream networks, those who are familiar with internet studies will immediately find that these are also the definitions of ``PV" and ``UV" of websites. Therefore, the online version of Kleiber's law, to exist, predicts that

\begin{equation}
\label{eq.1}
{PV_t} = A {UV_t}^{\theta}
\end{equation}

In which $A$ is a normalized coefficient. We argue that the exponent $\theta$ in Eq.1 not only shapes the growth dynamics of forums, but also provides a measure of the ``stickiness" of forums as an alternative to the average surfing length $L$, which is suggested in \cite{bucklin2002choice}. Using the indicator of $\theta$, we can easily separate ``sticky" forums from ``non-sticky" forums. In particular, if $\theta > 1$, we derive that $L_t = PV_t/UV_t \sim {PV_t}^{1-1/\theta > 0}$. It means that the average surfing length of users increases monotonically with forum size (or ``body mass"). In other words, users are more and more likely to be ``locked-in" in the forum during its growth. This is basically what we expect to see from a ``sticky" forum. On the contrary, if $\theta < 1$, users on average navigate less threads as the size of the forum increases, which is the property of a ``non-sticky" forum. An extra bonus of using $\theta$ as the indicator is that, $\theta = d(Log(PV_t)) / d(Log(UV_t))$ is a constant over time, whereas $L_t = PV_t/UV_t$ is not. Therefore, $\theta$ quantifies the ``stickiness" of forums as a stable, long-term property.

\subsection{\label{sec:3.2} The flow network expression of forums}

  \begin{figure*}[!ht]
    \centering
    \includegraphics[scale=0.6]{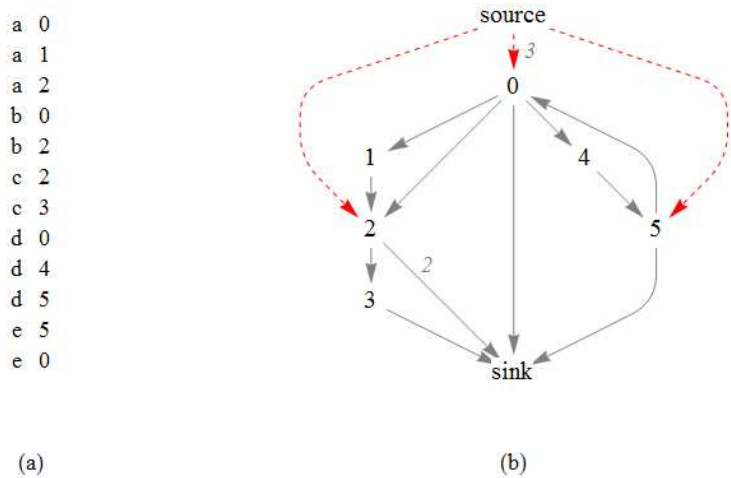}
    \caption{An example dataset and the clickstream network constructed from this dataset. In (b) the weights on edges are always $1$ unless otherwise specified by the number near the edge in gray color.}
    \label{fig.1}
  \end{figure*}

In the aforemention sections, we have discussed the forums as online flow systems. But in order to develop a framework that allows for quantitative analysis, we need a more explicit expression of the flow systems. This is how clickstream network comes in. Figure 1 shows the clickstream network constructed from a demo dataset, which contains the browsing histories of nine users. Each row corresponds to a single browsing activity. The first column denotes the cookies of users and the second column denotes the numeric IDs of visited threads. There is no duplicate record in the dataset. The example clickstream network is constructed as follows. For each record in the dataset, say, $[a, 0]$, if the next record has the same cookie $a$, e.g., $[a, 1]$, we add a clickstream from $0$ to $1$; otherwise, we create a clickstream from $0$ to the artificially added node ``sink". After all records are converted into clickstreams, we add a ``source" node to balance the network such that the in-flow (weighted in-degree) is equal to the out-flow (weighted out-degree) over all nodes \cite{higashi1986extended}. This balancing process demonstrates a very important principle, the conservation of attention (clickstreams). This principle requires that in-flow must equal out-flow both on the node level and on the network level. According to this principle, attention shouldn't appear or disappear out of nowhere. All the flow of attention circulating in networks comes from, and will eventually returns to, the ``environment" (the offline world or other forums).

On each node $i$, we defined $T_i$ as the clickstreams to $i$ and $D_i$ the clickstream dissipated by $i$. Base on the principle of attention conservation, $PV$ and $UV$ can be defined either on the node level or on the network level. On the node level, they are the the sums of $T_i$ and $D_{i}$ over all nodes in the network, respectively. On the network level, $PV$ is equal to the total weights of edges (before network balancing) and $UV$ is equal to the dissipation of the network, i.e., the in-flow of ``sink" or the out-flow of ``source" (which are equal to each other after network balancing). The equivalence of the two versions of definitions are non-trivial, because this allows us to explain the dynamics of network by the behavior of nodes, as to be shown in the next section.

At this point, we should pay attention to a major difference between biological/ecological and virtual flow systems. In biological and ecological flow networks, There is generally a ``root" node who is responsible for getting flow from the environment and supplying it to the rest of nodes, such as the roots of trees (who absorb water from the earth) or the producer in food webs (who gets energy from the Sun). As a consequence, the non-root nodes only dissipate and do not receive flow directly from the environment. flow to the environment. However, things are different in clickstream networks. There is no such a ``root" node. Actually, every node may receive flow directly from the environment. This is because the Web is, at least theoretically, a ``flat" world in the sense that users may enter into this world from any web page. As a result, we can calculate on each node both the in-flow from ``source" (which can be expressed as $I_{i}$) and the out-flow to ``sink" ($D_{i}$). To keep consistent with other flow network models \cite{west1997general,banavar1999size}, we define $UV$ as the sum of $D_{i}$ rather than the sum of $I_{i}$. In the last section we present a figure to briefly introduce the relationship between $I_{i}$ and $D_{i}$.

\subsection{\label{sec:3.3} The connection between growth and dissipation}

  \begin{figure*}[!ht]
    \centering
    \includegraphics[scale=0.7]{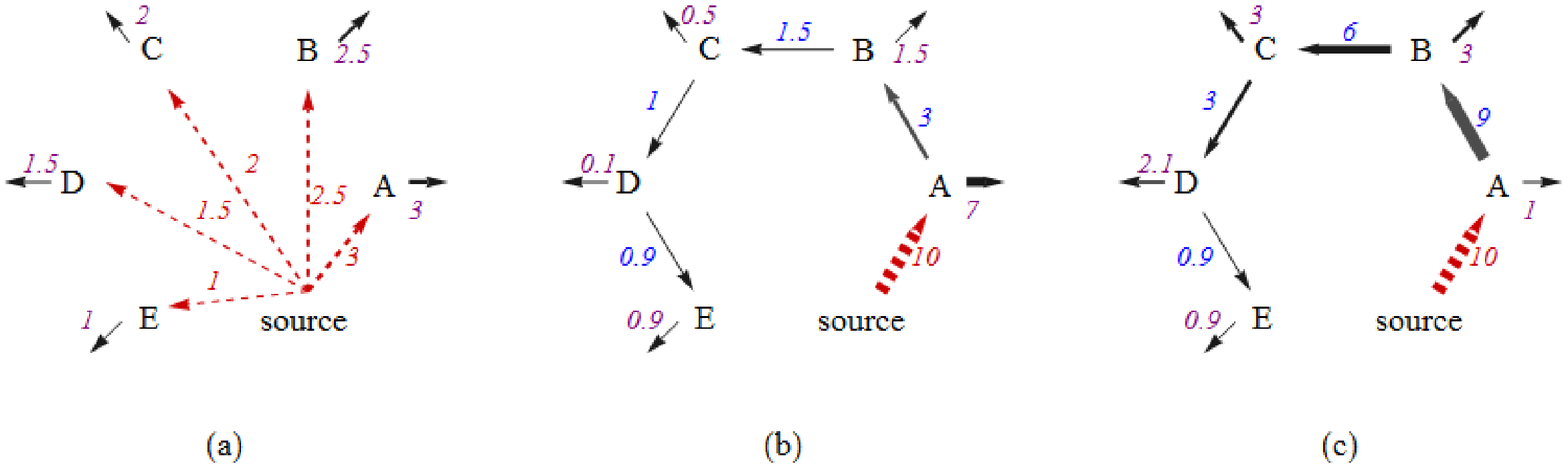}
    \caption{Three example clickstream networks of different topologies. (a) A star-like network in which the dissipation probability equals $100\%$. (b) A chain-like network in which the dissipation probability increases with the clickstreams to nodes. (c) A chain-like network in which the dissipation probability decreases with the clickstreams to nodes.
}
    \label{fig.2}
  \end{figure*}

In the last section, we discuss the definitions of the dissipated clickstream $D_i$ and the pass-through clickstream $T_i$. The dissipation law in ecology \cite{stravskraba1999ecosystems} predicts that

\begin{equation}
\label{eq.2}
{D_i} = B {T_i}^{\gamma}
\end{equation}

in which $B$ is a normalized coefficient and $\gamma$ is an exponent that reflects the efficiency of network dissipation. In Figure 2 we present three example flow networks to explain why $\gamma$ is a measure of dissipation efficiency and how it is related to $\theta$ in Eq.1.

First of all, let's consider two extreme topologies, the star-like (Figure 2a) and the chain-like (Figure 2b and 2c). In the star-like topology, all threads (nodes) receive clickstreams directly from the ``environment" and dissipate all clickstreams immediately; in the chain-like topology, all threads receive clickstreams sequentially from one another and dissipate a portion of the received clickstreams. For the convenience of comparison, we fix the $UV$ of all the three clickstream networks to be the same, i.e., 10. However, we find that the resulted $PV$ is larger in the chain-like networks (10+3+1.5+1+0.9 = 16.4 in (b) and 10+9+6+3+0.9 = 28.9 in (c) ) than in the star-like network (3+2.5+1.5+1 = 10 in (c)). This is because by transporting clickstreams between threads instead of dissipating clickstreams immediately, the network increases its storage capacity of clickstreams, i.e., the ``body mass". To understand this interesting phenomenon, one can consider how a clown plays balls. A clown can barely hold more than two balls if he just grasp them in his hands, but he can easily maintain a circulation of many balls by throwing up and passes the balls from one hand to the other when they fall down. It is in exactly the same way that clickstream transportation increases the total amount of clickstreams ``hold" by a network.

So how dissipation efficiency $\gamma$ is related to transportation? We find that the smaller $\gamma$ ``delays" the dissipation of clickstreams and thus increases the amount of clickstreams transported in the network. This finding is demonstrated by the comparison between Figure 2b and 2c. To facilitate the comparison, we define the log-out probability of users $P_i = D_i/T_i \sim {T_i}^{\gamma-1}$. Therefore, in networks with $\gamma >1$ $P_i$ increases with the clickstreams to nodes and otherwise it decreases with the clickstreams. We calculate that $P_{ib}=\{70\%, 50\%, 30\%, 10\%\}$ from node $A$ to $D$ in Figure 2b and $P_{ic}=\{10\%, 30\%, 70\%, 50\%\}$ in Figure 2c (for the convenience of the comparison, we ignore the behavior of node E, whose clickstreams are very small compared to other nodes). As the pass-though clickstreams decrease monotonously from $A$ to $D$, it is easy to derive that $\gamma_b >1> \gamma_c$. As we have calculated that $UV_b=UV_c$ and $PV_b<PV_c$, we have $\theta_b<\theta_c$. From this simple deduction, we conjecture that $\gamma$ and $\theta$ are negatively correlated. In fact, it is reasonable to expect this negative correlation in clickstream networks of different topologies. Because the lower $\gamma$ always forces large nodes to transport clickstreams to other nodes instead of dissipating them to the environment. This process increases the pass-through flow of the down-stream nodes of these large nodes, and also the downstream nodes of the downstream nodes, and so on. This process continues, remaining more and more flow within the network, until the rest flow is eventually dissipated to the environment at the boundary of the diffusion area. It is obviously that such a flow diffusion process will always increase the ``stickiness" of a flow network no matter what shape it has.

\section{Results}

In data analysis, we use the log file of Baidu Tieba to construct hourly-based clickstream networks. Among the millions of forums in the entire Baidu Tieba system, we select the top 30,000 forums, whose size (the averaged daily page views in two months) varies from hundreds to millions. For each forum, we construct 1,440 successive hourly-based clickstream networks using the historical browsing data in two months (from Feb. 27, 2013 to Apr. 27, 2013). We calculate $UV_t$ and $PV_t$ of these networks to derive $\theta$. In analyzing the dissipation behavior of nodes, we randomly select a day (Apr. 24, 2013) and construct 24 successive hourly networks. In fact, to estimate $\gamma$, we just need one hourly clickstream network. The reason to include 24 networks is to test whether the estimation of $\gamma$ is robust over time. In analyzing the relationship between $\gamma$ and $\theta$ we use the averaged value of $\gamma$ in 24 hours. In the analysis of the scaling relationships expressed in Eq.1 and Eq.2, we always use the ordinary least square regression in log-log plots to estimate the scaling exponent \cite{brown2004toward}.

\subsection{\label{sec:4.1} The allometric growth of forums }

  \begin{figure*}[!ht]
    \centering
    \includegraphics[scale=0.5]{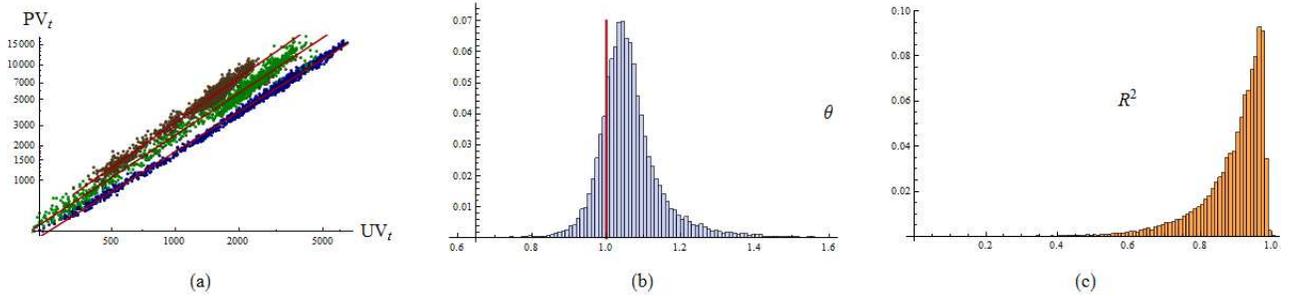}
    \caption{(a) The scaling relationships between $UV_t$ and $PV_t$ across three forums in 1,440 hours. Each data point correspond to a pair of $UV_t$ and $PV_t$. Data points of different forums are shown in different colors. The values of $\theta$ are $1.15$ (blue points), $1.21$ (green points), and $1.29$ (brown points) in the three forums, respectively. (b) The distribution of $\theta$ of the $29,993$ forums (the data of the rest 7 forums are not enough to support validated fitting). The mean value is $1.06$ and the standard deviation (SD) is $0.10$. (c) The distribution of $R^2$ in fitting $\theta$. The mean value is $0.89$ and the SD is $0.10$.}
    \label{fig.3}
  \end{figure*}

Figure 3a shows how Eq.1 shapes the growth dynamics of three different forums during the studied period. We find that this strong regularity holds for most of the forums: more then $86\%$ of forums have a $R^2 > 0.8$ in the fitting of Eq.1. This finding supports the assumption that all real-world flow systems are governed by the same underlying mechanics and thus exhibit similar regularities \cite{bejan2011constructal}. In fact, we have extended this assumption to include both of real-world and virtual flow systems. It is very inspiring to find that human attention, after being quantified as clickstreams, confirms to the physical laws observed in natural systems.

In Kleiebr's law, the ``body mass" scales to ``energy consumption" with an exponent $4/3 \approx 1.33$ \cite{west1997general}. But the exponent observed in our data is generally smaller than this value. The mean value of $\theta$ is $1.06$ and the standard deviation ($SD$) is $0.10$ (Figure 3b). As shown by Figure 3b, the shape of the distribution is slightly asymmetrical; it skews towards the right hand side of the $x$ axis beyond the point of (x=1,y=0). In particular, $82\%$ of the forums has a $\theta >1$. These results suggest that most forums are ``sticky", in the sense that users are more likely to be attracted and remained in the forums when the forum size grows. However, by comparing $\theta$ between virtual and real flow systems, we know that the stickiness of forums to user's attention is less than that of biological organisms to water and nutrient: the latter are able to remain the flow in the systems for a longer time before dissipating them to the environment. How can the forums learn from biological organisms ?  This is an interesting topic beyond the scope of the current research, but worth being looked into in future studies

\subsection{\label{sec:4.2}The scaling of clickstream dissipation}

  \begin{figure*}[!ht]
    \centering
    \includegraphics[scale=0.5]{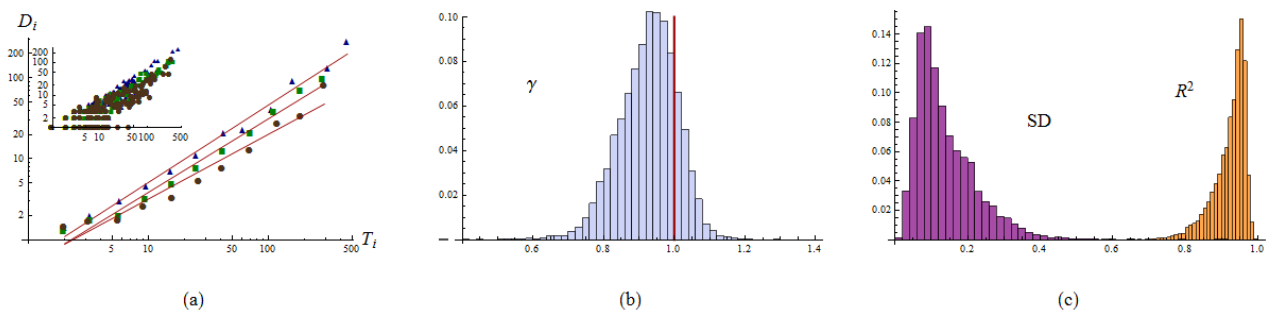}
    \caption{(a) The scaling relationships between $T_i$ and $D_i$ across forums in three hourly networks. These three forums are the same as the forums presented in Figure 3a. The color scheme of these data points is the same as that of Figure 3a. The value of $\gamma$ are $0.96$ (blue triangles), $0.90$ (green squares), and $0.80$ (brown circles) for the three forums, respectively. (b) The distribution of the averaged value of $\gamma$ over 24 hours across the $6,877$ forums. The mean value of the distribution is $0.93$ and the $SD$ is $0.08$. (c) The distribution of the $SD$ of $\gamma$ over 24 hours (purple bars) and the averaged $R^2$ in fitting $\gamma$ (orange bars). The mean and $SD$ of the two distributions are $0.14$ and $0.09$, and $0.92$ and $0.05$, respectively.
}
    \label{fig.4}
  \end{figure*}

We find that the law of dissipation (Eq.2) is a robust pattern holds for most of the studied forums: more then $98\%$ of forums have a $R^2 > 0.8$ in the fitting of Eq.2. Meanwhile, the value of $\gamma$ estimated from hourly networks is a stable quantity over time (the $SD$ of $\gamma$s in 24 hours is $0.14$). The mean and $SD$ of the $\gamma$ distribution is $0.93$ and $0.08$, respectively. On the contrary of the $\theta$ distribution (Figure 3b), the distribution of $\gamma$ skews towards the left hand side of the $x$ axis beyond the point of (x=1,y=0): $82\%$ of forums has a $\gamma <1$. According to aforementioned discussions, this means that most of the studied forums have a low dissipative efficiency, i.e., the log-out probability of users decreases with the clickstreams to threads.

This findings help us understand the usage pattern of Tieba forums. In browsing threads, users are more likely to log out from non-popular threads than popular threads. There are various factors that may contribute to this phenomenon, but we conjecture that the reverse-time displaying order of threads, together with the ``bumping" mechanism, is probably one of the major reasons. All the forums in Tieba system sort threads in reverse time (which is the time of the latest comment) order and display them in sequential pages. ``Bump" describes an action (e.g., posting) taken by a user such that a particular thread is returned to the top of the thread list. Some users may even post a message with only the term ``bump" to show that they are bumping to make sure that more users will see the thread. Popular threads benefit greatly from ``bumping" are always displayed on the first page. As these forums are all interest groups with specific topic, such as ``the Fans of Lady Gaga" or ``Everything about Star Wars", visitors generally share common interest. Instead of selective reading (which is more common in platforms such as news aggregators), they usually browse the threads in the default displaying order from the top to the bottom and from the first page to the other pages. As a result, before users getting tired, they have read the most popular threads. That is why we observe that users are more likely to stop browsing at non-popular threads.

\subsection{\label{sec:4.3}The negative correlation between $\gamma$ and $\theta$}

  \begin{figure*}[!ht]
    \centering
    \includegraphics[scale=0.6]{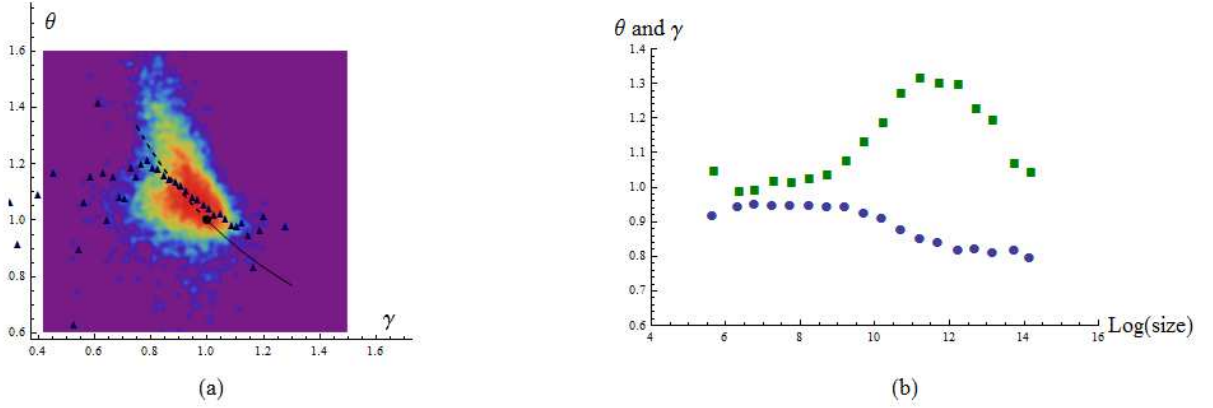}
    \caption{The negative correlation between $\gamma$ and $\theta$ (a) and the change of $\gamma$ and $\theta$ with forum size (b). In (a) We plot both of the ``binned" data (blue triangles) and the original data (heat map). In the heat map, the warmer color means that the distribution of the data points is more dense. In plotting the ``binned" data, we calculated the average of $x$ and $y$ values in intervals uniformly selected from the $x$ range. This technique is frequently used to eliminate the noise in data \cite{newman2005power}.
}
    \label{fig.4}
  \end{figure*}

We find that the exponent of the dissipation law, $\gamma$, has a negative correlation with $\theta$ as expected. To understand this finding more deeply, we conduct a simple mathematical analysis. As introduced in Section 3.2, $UV$ is equal to the sum of $D_i$ and $PV$ is equal to the sum of $T_i$. Thus, we have:

\begin{equation}
\label{eq.3}
{PV_t} = A {UV_t}^{\theta} \rightarrow {UV_t} = {(PV_t/A)}^{1/\theta} = {( \sum T_i)}^ {1/\theta}/ A^{1/\theta}
\end{equation}

\begin{equation}
\label{eq.4}
{D_i} = B {T_i}^{\gamma} \rightarrow {UV_t} = { \sum D_i} = { \sum B T_i^ {\gamma}} =  B { \sum T_i^ {\gamma}}
\end{equation}

Joining Eq.3 and Eq.4, we have :

\begin{equation}
\label{eq.5}
{( \sum T_i)}^ {1/\theta}/\sum {T_i}^ {\gamma} = A^{1/\theta} B
\end{equation}

In data analysis, we find that $0<A<1$, $0<B<1$ and $\theta>0$, thus $0<A^{1/\theta} B<1$. Using this condition, we derive that

\begin{equation}
\label{eq.6}
{( \sum T_i)}^ {1/\theta} <  \sum {T_i}^ {\gamma} < {( \sum T_i)}^ {\gamma} \rightarrow  \theta > 1/\gamma
\end{equation}

Eq. 6 holds only when $\gamma>1$. In other words, if $\gamma>1$, $1/\gamma$ is the lower boundary of $\theta$. To those cases where $\gamma<1$, unfortunately we can not give further analytical conclusions. We only know that when $\gamma$ is smaller than, but close to $1$, $\theta$ approaches $1/\gamma$. To validate our derivations, we $1/\gamma$ in black line (which is separated into two parts, the dashed one and solid one, by the point [1,1] ) in Figure 5a . We find that most of the binned data points are located above the boundary denoted by the solid line, supporting our derivation. By analyze the binned data and the original data, we find that the estimation of $\gamma$ is reliable in the range [0.8,1.1], which correspond to the values of $\gamma$ in [1.0, 1.2]. Beyond this scope, the estimations of the two parameters vary sharply due to a lack of data.

To summarize, the reverse-time displaying order of threads, combined with the ``bumping" mechanism, seems to decrease the dissipation efficiency $\gamma$ and thus increase the ``stickiness" $\theta$ of forums. Is this the reason why the reverse-time displaying order is so frequently used in forums and other online communities? This is an interesting question worth further exploration.

At this stage, a natural question to ask is whether $\gamma$ and $\theta$ is affected by the forum size. To answer this question, we plot these two quantities against forum size as shown in Figure 5b. We find that when the forum size approximates $10^5$, $\gamma$ reaches its minimum value and $\theta$ reaches its maximum value.

\section{Discussion and Conclusion}

\subsection{\label{sec:5.1}The asymmetric between imported and dissipated clickstreams }

  \begin{figure*}[!ht]
    \centering
    \includegraphics[scale=0.6]{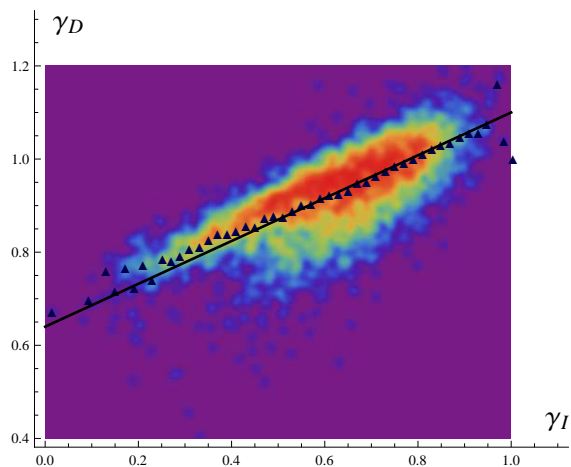}
    \caption{The linear relationship between $\gamma_{D}$ and $\gamma_{I}$. We plot both of the ``binned" data (blue triangles) and the original data (heat map). In the heat map, the warmer color means that the distribution of the data points is more dense. The slope of the regression line fitted from the binned data is $0.46$.
}
    \label{fig.6}
  \end{figure*}

As discussed in Section 3.2, we can calculate on each node both the in-flow from ``source" ($I_{i}$) and the out-flow to ``sink" ($D_{i}$) in a clickstream network. The parameter $\gamma$ in Eq.2 is actually $\gamma_{D}$. By replacing $D_{i}$ in Eq.2 with $I_{i}$, we can estimate the value of $\gamma_{I}$.  $\gamma_{D}$ describes the dissipation behavior of nodes and $\gamma_{I}$ describes the flow importing behavior of nodes. In most of flow networks in the real-world, there is only one ``root-node" that takes the responsibility of importing flow. But in clickstream networks this job is assigned distributively to many nodes. Figure 6 shows that $\gamma_{D}$ changes with $\gamma_{I}$ linearly with a coefficient that equals $0.46$. Both of the two parameters are smaller than $1$, this means that in the network, large nodes always derive flow from,and return flow back to small nodes instead of exchange flow with the environment. This finding confirmed our assumption on the ``chain-like" topology of the studied clickstream networks.

\subsection{\label{sec:5.2}The displaying order of threads and the stickiness of forums}

As mentioned in Section 4.2, the display order is a key factor that affects the log-out probability of users as well as the stickiness of forums. To understand the effect of different displaying schemes, one can consider the following metaphor.

Webmasters and users are like two players who are playing the game of porker. The rule is that the two players draw a fixed number of cards from randomly shuffled decks and for each round they present one card to compare by rank. If the webmaster's card is lower than that of the user's, he loses and the user will quit the game. The goal of the webmaster, therefore, is to take a playing strategy such that he can play more rounds. In this metaphor, the webmaster's strategy represents the displaying mechanism of threads, the users' strategy represents his preference, and the rule that the webmaster's card should be higher in rank to continue the game corresponds to user's navigation decision: in each step of the navigation, if the visited thread provides a utility higher than the expectation of the user, he will continue the navigation; otherwise, he is very likely to log out from the forum.

Traditionally, the webmaster's strategy is fixed, whereas user's strategy varies across individuals. There are basically three strategies: 1) to present cards randomly; 2) to present cards in increasing order; and 3) to present cards in decreasing order. As the strategy of users varies, we assume that the average user takes the randomly strategy. Now it is obviously that the last strategy, that is, to present cards in decreasing order, will maximize the successive winning rounds and should be preferred by the webmaster. In other words, the preferable displaying strategy of forums is to show thread in decreasing popularity ($PV$).

The strategy used by Tieba system is the combination of the reverse-time displaying order and the ``bumping" mechanism. This appears to be different from the decreasing popularity strategy, but as discussed, they lead to the similar result. Now we have a more intuitive understanding on how the displaying mechanism of Tieba forums leads to user's logging out at non-popular threads (whose utility are smaller than the expected utility of users) and thus increase the ``stickiness" of forums. Please note that in the above metaphor we assume that the webmasters' strategy is fixed. The introduction of other mechanisms, such as personalized recommendation \cite{shepitsen2008personalized}, will dramatically change the rule of the game. Actually, personalized recommendation is like a cheating method that make the webmaster know the cards selected by the users in advance, and present cards accordingly.

\subsection{\label{sec:5.3}Summary }

Websites, by its very nature, are the consumers of collective attention and the producers of information \cite{huberman2003laws}. We suggest that the comparison of websites with living organisms are not just qualitative metaphors, but also provides quantitative insights into the understanding of websites usage. In this study, we find substantial evidence that websites confirm to the same constraints that also shape the evolve of natural systems.

In particular, we discover the online version of Kleibers' law, that is, the scaling relationship between $UV_t$ and $PV_t$ in the temporal evolution of forums. Further, we show that the allometric exponent $\theta$, which is an indicator for the ``stickiness" of forums in attracting users, is determined by the metabolism of clickstream networks. The lower the dissipation efficiency $\gamma$ is, the larger the $\theta$ would be. Interesting, there seems to be an optimized scale of forums at around $10^5$ daily $PV$s that minimizes $\gamma$ and maximizes $\gamma$.

As suggested by Bettencourt et al. \cite{bettencourt2007growth}, the allometric growth is a very general relationship between variables in the evolution of complex systems. In particular, they show that cities are extensions of biological entities, in the sense that they satisfy the same allometric functions \cite{bettencourt2007growth,brown2004toward}. Our study extends their findings in offline social systems to online social systems. We agree with Bettencourt et al. that the scaling relationship does not need to be restricted between body size and energy consumption, but is also applicable to other variables. For example, the recently found ``densification" pattern in the growth of online networks \cite{leskovec2007graph}, together with the scaling relationships discussed in \cite{cattuto2009collective,tessone2011sustainable,wu2011accelerating,henderson2001modelling,leskovec2007graph}, are probably different expressions of the ``allometric growth" of online flow networks.

Our findings is relevant to the Web development in many aspects. In particular, the present method to predict the long-term trend of clicks on an online community by analyzing user behavior within a short time period is useful in click prediction and other areas of computational advertisement \cite{kim2004clickstream}. To evaluate the ``stickiness" $\theta$ of forums, instead of monitoring the $PV$ and $UV$ in months, one simply needs to trace the log-out probability $P_{i}$ of users on a random sample of threads in a hour, and examines the correlation between $P_{i}$ and the clickstreams to the thread $T_{i}$. The direction of this correlation reflects the sticky/non-sticky property (negative for sticky and positive for non-sticky) and the magnitude of the correlation indicates the level of sticky/non-sticky. Another application of $\theta$ is to integrate it as a novel feature in to the recommendation of interest-groups \cite{fu2000mining}.

\begin{acknowledgements}

\end{acknowledgements}

\bibliographystyle{unsrt}
\bibliography{metabolism}

\providecommand{\noopsort}[1]{}\providecommand{\singleletter}[1]{#1}%
\begin{thebibliography}{10}

\bibitem{o2007web}
Tim O'reilly.
\newblock What is web 2.0: Design patterns and business models for the next
  generation of software.
\newblock {\em Communications \& strategies}, (1):17, 2007.

\bibitem{top2012500}
Alexa Top.
\newblock 500 global sites, 2011.
\newblock {\em URL http://www. alexa. com/topsites}, 2012.

\bibitem{bodendorf2009detecting}
Freimut Bodendorf and Carolin Kaiser.
\newblock Detecting opinion leaders and trends in online social networks.
\newblock In {\em Proceedings of the 2nd ACM workshop on Social web search and
  mining}, pages 65--68. ACM, 2009.

\bibitem{cammaerts2005online}
Bart Cammaerts and Leo~van Audenhove.
\newblock Online political debate, unbounded citizenship, and the problematic
  nature of a transnational public sphere.
\newblock {\em Political Communication}, 22(2):179--196, 2005.

\bibitem{zhang2007expertise}
Jun Zhang, Mark~S Ackerman, and Lada Adamic.
\newblock Expertise networks in online communities: structure and algorithms.
\newblock In {\em Proceedings of the 16th international conference on World
  Wide Web}, pages 221--230. ACM, 2007.

\bibitem{abbasi2005applying}
Ahmed Abbasi and Hsinchun Chen.
\newblock Applying authorship analysis to extremist-group web forum messages.
\newblock {\em Intelligent Systems, IEEE}, 20(5):67--75, 2005.

\bibitem{benevenuto2009characterizing}
Fabr{\'\i}cio Benevenuto, Tiago Rodrigues, Meeyoung Cha, and Virg{\'\i}lio
  Almeida.
\newblock Characterizing user behavior in online social networks.
\newblock In {\em Proceedings of the 9th ACM SIGCOMM conference on Internet
  measurement conference}, pages 49--62. ACM, 2009.

\bibitem{yu2010analyzing}
Jiefei Yu, Yanqing Hu, Min Yu, and Zengru Di.
\newblock Analyzing netizens’ view and reply behaviors on the forum.
\newblock {\em Physica A: Statistical Mechanics and its Applications},
  389(16):3267--3273, 2010.

\bibitem{west1997general}
Geoffrey~B West, James~H Brown, and Brian~J Enquist.
\newblock A general model for the origin of allometric scaling laws in biology.
\newblock {\em Science}, 276(5309):122--126, 1997.

\bibitem{brown2004toward}
James~H Brown, James~F Gillooly, Andrew~P Allen, Van~M Savage, and Geoffrey~B
  West.
\newblock Toward a metabolic theory of ecology.
\newblock {\em Ecology}, 85(7):1771--1789, 2004.

\bibitem{zhang2010scaling}
Jiang Zhang and Liangpeng Guo.
\newblock Scaling behaviors of weighted food webs as energy transportation
  networks.
\newblock {\em Journal of Theoretical Biology}, 264(3):760--770, 2010.

\bibitem{bucklin2002choice}
Randolph~E Bucklin, James~M Lattin, Asim Ansari, Sunil Gupta, David Bell,
  Eloise Coupey, John~DC Little, Carl Mela, Alan Montgomery, and Joel Steckel.
\newblock Choice and the internet: From clickstream to research stream.
\newblock {\em Marketing Letters}, 13(3):245--258, 2002.

\bibitem{stravskraba1999ecosystems}
Milan Stra{\v{s}}kraba, Sven~E J{\o}rgensen, and Bernard~C Patten.
\newblock Ecosystems emerging: 2. dissipation.
\newblock {\em Ecological Modelling}, 117(1):3--39, 1999.

\bibitem{zhang2013allometry}
Jiang Zhang and Lingfei Wu.
\newblock Allometry and dissipation of ecological flow networks.
\newblock {\em arXiv preprint arXiv:1302.5803}, 2013.

\bibitem{west1999fourth}
Geoffrey~B West, James~H Brown, and Brian~J Enquist.
\newblock The fourth dimension of life: fractal geometry and allometric scaling
  of organisms.
\newblock {\em Science}, 284(5420):1677--1679, 1999.

\bibitem{banavar1999size}
Jayanth~R Banavar, Amos Maritan, and Andrea Rinaldo.
\newblock Size and form in efficient transportation networks.
\newblock {\em Nature}, 399(6732):130--132, 1999.

\bibitem{garlaschelli2003universal}
Diego Garlaschelli, Guido Caldarelli, and Luciano Pietronero.
\newblock Universal scaling relations in food webs.
\newblock {\em Nature}, 423(6936):165--168, 2003.

\bibitem{cheng2010personalized}
Haibin Cheng and Erick Cant{\'u}-Paz.
\newblock Personalized click prediction in sponsored search.
\newblock In {\em Proceedings of the third ACM international conference on Web
  search and data mining}, pages 351--360. ACM, 2010.

\bibitem{fu2000mining}
Xiaobin Fu, Jay Budzik, and Kristian~J Hammond.
\newblock Mining navigation history for recommendation.
\newblock In {\em Proceedings of the 5th international conference on
  Intelligent user interfaces}, pages 106--112. ACM, 2000.

\bibitem{huberman2009social}
Bernardo~A Huberman.
\newblock Social attention in the age of the web.
\newblock {\em Working together or apart: Promoting the next generation of
  digital scholarship}, page~62, 2009.

\bibitem{cooley2000web}
Robert~Walker Cooley.
\newblock {\em Web usage mining: discovery and application of interesting
  patterns from web data}.
\newblock PhD thesis, University of Minnesota, 2000.

\bibitem{huberman1998strong}
Bernardo~A Huberman, Peter~LT Pirolli, James~E Pitkow, and Rajan~M Lukose.
\newblock Strong regularities in world wide web surfing.
\newblock {\em Science}, 280(5360):95--97, 1998.

\bibitem{johnson2003cognitive}
Eric~J Johnson, Steven Bellman, and Gerald~L Lohse.
\newblock Cognitive lock-in and the power law of practice.
\newblock {\em Journal of Marketing}, pages 62--75, 2003.

\bibitem{bucklin2003model}
Randolph~E Bucklin and Catarina Sismeiro.
\newblock A model of web site browsing behavior estimated on clickstream data.
\newblock {\em Journal of Marketing Research}, pages 249--267, 2003.

\bibitem{bollen2009clickstream}
Johan Bollen, Herbert Van~de Sompel, Aric Hagberg, Luis Bettencourt, Ryan
  Chute, Marko~A Rodriguez, and Lyudmila Balakireva.
\newblock Clickstream data yields high-resolution maps of science.
\newblock {\em PLoS One}, 4(3):e4803, 2009.

\bibitem{huberman2009crowdsourcing}
Bernardo~A Huberman, Daniel~M Romero, and Fang Wu.
\newblock Crowdsourcing, attention and productivity.
\newblock {\em Journal of Information Science}, 35(6):758--765, 2009.

\bibitem{lerman2010information}
Kristina Lerman and Rumi Ghosh.
\newblock Information contagion: An empirical study of the spread of news on
  digg and twitter social networks.
\newblock {\em ICWSM}, 10:90--97, 2010.

\bibitem{wu2007novelty}
Fang Wu and Bernardo~A Huberman.
\newblock Novelty and collective attention.
\newblock {\em Proceedings of the National Academy of Sciences},
  104(45):17599--17601, 2007.

\bibitem{cattuto2007semiotic}
Ciro Cattuto, Vittorio Loreto, and Luciano Pietronero.
\newblock Semiotic dynamics and collaborative tagging.
\newblock {\em Proceedings of the National Academy of Sciences},
  104(5):1461--1464, 2007.

\bibitem{wu2009feedback}
Fang Wu, Dennis~M Wilkinson, and Bernardo~A Huberman.
\newblock Feedback loops of attention in peer production.
\newblock In {\em Computational Science and Engineering, 2009. CSE'09.
  International Conference on}, volume~4, pages 409--415. IEEE, 2009.

\bibitem{bettencourt2007growth}
Lu{\'\i}s~MA Bettencourt, Jos{\'e} Lobo, Dirk Helbing, Christian K{\"u}hnert,
  and Geoffrey~B West.
\newblock Growth, innovation, scaling, and the pace of life in cities.
\newblock {\em Proceedings of the National Academy of Sciences},
  104(17):7301--7306, 2007.

\bibitem{cattuto2009collective}
Ciro Cattuto, Alain Barrat, Andrea Baldassarri, Gregory Schehr, and Vittorio
  Loreto.
\newblock Collective dynamics of social annotation.
\newblock {\em Proceedings of the National Academy of Sciences},
  106(26):10511--10515, 2009.

\bibitem{tessone2011sustainable}
Claudio~J Tessone, Markus~M Geipel, and Frank Schweitzer.
\newblock Sustainable growth in complex networks.
\newblock {\em EPL (Europhysics Letters)}, 96(5):58005, 2011.

\bibitem{wu2011accelerating}
Lingfei Wu and Jiang Zhang.
\newblock Accelerating growth and size-dependent distribution of human online
  activities.
\newblock {\em Physical Review E}, 84(2):026113, 2011.

\bibitem{henderson2001modelling}
Tristan Henderson and Saleem Bhatti.
\newblock Modelling user behaviour in networked games.
\newblock In {\em Proceedings of the ninth ACM international conference on
  Multimedia}, pages 212--220. ACM, 2001.

\bibitem{leskovec2007graph}
Jure Leskovec, Jon Kleinberg, and Christos Faloutsos.
\newblock Graph evolution: Densification and shrinking diameters.
\newblock {\em ACM Transactions on Knowledge Discovery from Data (TKDD)},
  1(1):2, 2007.

\bibitem{higashi1986extended}
Masahiko Higashi.
\newblock Extended input-output flow analysis of ecosystems.
\newblock {\em Ecological Modelling}, 32(1):137--147, 1986.

\bibitem{bejan2011constructal}
Adrian Bejan and Sylvie Lorente.
\newblock The constructal law and the evolution of design in nature.
\newblock {\em Physics of Life Reviews}, 8(3):209--240, 2011.

\bibitem{newman2005power}
Mark~EJ Newman.
\newblock Power laws, pareto distributions and zipf's law.
\newblock {\em Contemporary physics}, 46(5):323--351, 2005.

\bibitem{shepitsen2008personalized}
Andriy Shepitsen, Jonathan Gemmell, Bamshad Mobasher, and Robin Burke.
\newblock Personalized recommendation in social tagging systems using
  hierarchical clustering.
\newblock In {\em Proceedings of the 2008 ACM conference on Recommender
  systems}, pages 259--266. ACM, 2008.

\bibitem{huberman2003laws}
Bernardo~A Huberman.
\newblock {\em The laws of the Web: Patterns in the ecology of information}.
\newblock MIT Press, 2003.

\bibitem{kim2004clickstream}
Dong-Ho Kim, Vijayalakshmi Atluri, Michael Bieber, Nabil Adam, and Yelena
  Yesha.
\newblock A clickstream-based collaborative filtering personalization model:
  towards a better performance.
\newblock In {\em Proceedings of the 6th annual ACM international workshop on
  Web information and data management}, pages 88--95. ACM, 2004.

\end{thebibliography}

\end{document}